\documentclass[%
 reprint,
superscriptaddress,
 amsmath,amssymb,
 aps, prb,
floatfix,
]{revtex4-2}

\usepackage{graphicx} 
\usepackage{dcolumn} 
\usepackage{sidecap}
\usepackage{float}
\usepackage{physics}
\usepackage{bm} 
\usepackage[svgnames]{xcolor}
\usepackage[bookmarks=false,linkcolor=blue,urlcolor=blue,colorlinks,citecolor=blue]{hyperref}
\usepackage{lipsum}

\begin{document}

\title{Chiral and topological superconductivity in isospin polarized multilayer graphene}

\author{Max Geier}
\affiliation{Department of Physics, Massachusetts Institute of Technology, Cambridge, MA 02139, USA}
\author{Margarita Davydova}
\affiliation{Department of Physics, Massachusetts Institute of Technology, Cambridge, MA 02139, USA}
\affiliation{Walter Burke Institute for Theoretical Physics and Institute for Quantum Information and Matter,
California Institute of Technology, Pasadena, CA 91125, USA}
\author{Liang Fu}
\affiliation{Department of Physics, Massachusetts Institute of Technology, Cambridge, MA 02139, USA}

\date{\today} 

\begin{abstract}
A microscopic mechanism for chiral $p$-wavesuperconductivity from Coulomb repulsion is proposed for spin- and valley-polarized state of rhombohedral multilayer graphene. 
The superconducting instability arises when strong Thomas-Fermi screening of the Coulomb potential allows Friedel oscillations to take over -- leading to an effective attraction on length scales below the Fermi wavelength.
The superconducting critical temperature is largest at low density below a Lifshitz transition to an annular Fermi sea, where the additional pocket strongly enhances Thomas-Fermi screening.
The Lifshitz transition also marks a topological phase transition from a trivial to a topological superconducting phase hosting Majorana fermions. 
The chirality of the superconducting order parameter is selected by the chirality of the valley-polarized Bloch electrons. 
Our results are in reasonable agreement with observations in a recent experiment on tetralayer graphene [Han, T., Lu, Z., Hadjri, Z. {\it et al.}, {\it Nature} (2025) \cite{Han2024Aug}]. 
\end{abstract}

\maketitle

Chiral superconductivity, characterized by spontaneous time-reversal symmetry breaking and finite-angular momentum Cooper pairing \cite{Kallin2016Apr}, is a long-sought quantum phase of matter with unusual superconducting and magnetic properties. Interest in chiral superconductors is further fueled by their potential for hosting topological phases and Majorana fermions \cite{Read2000Apr,
Sato2017May
}. While previous material candidates, such as ${\rm Sr}_2{\rm RuO}_4$ \cite{Maeno1994Dec} and ${\rm UTe}_2$ \cite{Aoki2019Mar,Jiao2020Mar,Aoki2022Apr}, showed initial signs of chiral superconductivity, recent experiments strongly suggest single-component superconducting order parameters that are non-chiral \cite{Kallin2009Mar,Mackenzie2017Jul,Romer2019Dec,Kivelson2020Jun,Roising2022Nov,Ajeesh2023Oct,Azari2023Nov,Andersen2024May}.

Very recently, signatures of chiral superconductivity have been observed in rhombohedral-stacked tetralayer graphene under electron doping \cite{Han2024Aug}. While superconductivity has been previously discovered and intensively studied in crystalline trilayer \cite{Zhou2021Oct,Chou2021Oct,Ghazaryan2021Dec,Li2021Dec,You2022Apr,Chatterjee2022Oct,Ghazaryan2023Mar,Jimeno-Pozo2023Apr,Qin2023Apr,Pantaleon2023May,Li2023Jul,Dong2023Oct,Dong2024Jun,murshedPRB2025}
and bilayer graphene \cite{Zhou2022Jan,Zhang2023Jan,Li2024Jul,Cea2022Feb,Chou2022Mar,Dong2023May}
, the newly found superconducting state in tetralayer graphene at low density is remarkably distinctive in that they exhibit large spontaneous anomalous Hall effect above $T_c$ and magnetic hysteresis in resistance below $T_c$. These observations demonstrate time-reversal-breaking superconductivity in a pure carbon system. Its pairing symmetry and pairing mechanism are open questions for investigation.   

A key feature of rhombohedral multilayer graphene is the flat band dispersion near $K$ and $K'$ points leading to strong correlation effect \cite{Ghazaryan2021Dec,Ghazaryan2023Mar}. As a result, spin and valley isospin symmetry breaking occurs at low temperature, giving rise to half and quarter metal phases \cite{Ghazaryan2023Mar}. Interestingly, the superconducting state in tetralayer graphene at low density borders the quarter metal and their phase boundary shows no or little change with the applied magnetic field, indicating that the superconducting state is likely fully spin and valley polarized \cite{Han2024Aug}. Thus, tetralayer graphene provides a rare opportunity for investigating Cooper pairing of single flavor electrons in a solid state platform.

\begin{figure}[!t]
\includegraphics[width= \columnwidth]{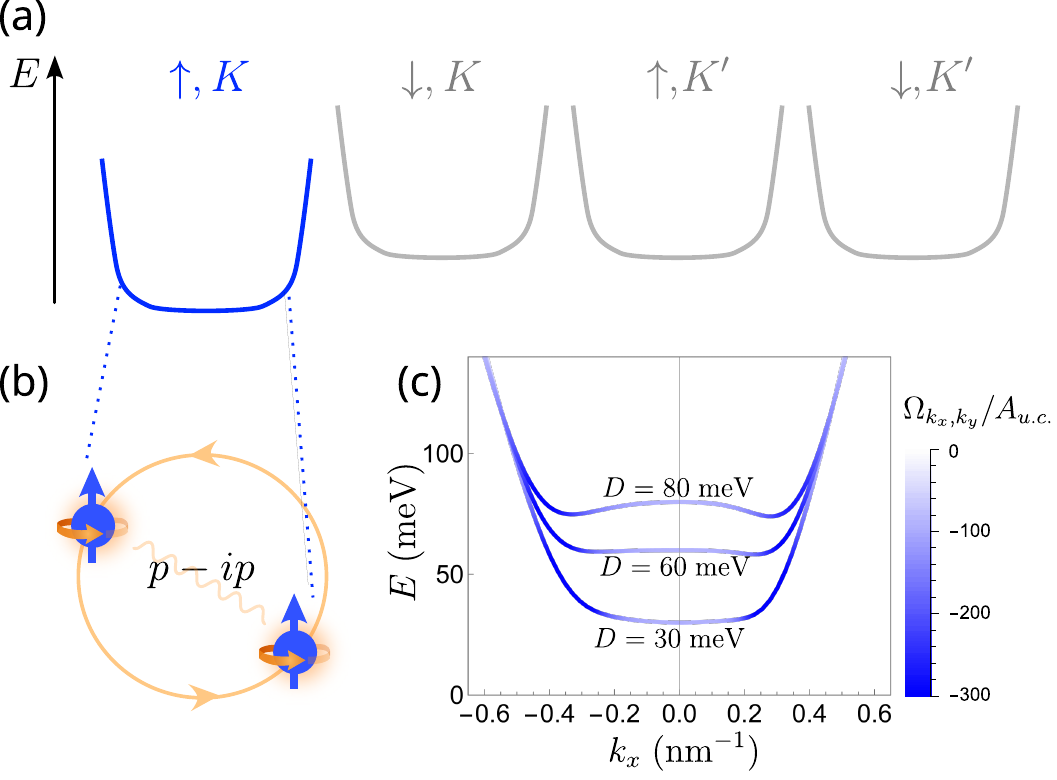}
\caption{
(a) Schematic of a chiral superconducting pairing appearing on top of an isospin-polarized quarter-metal state in multilayer graphene. 
(b) The chirality of the superconducting state is selected by the orbital magnetic moment of the electrons in the polarized valley. 
(c) Dispersion of ABCA graphene around the $K$ valley \cite{Ghazaryan2021Dec,Ghazaryan2023Mar} for various electric potential differences $D$ between top and bottom layers. The line color indicates the Berry curvature $\Omega_{k_x,k_y}$.}
\label{fig:1} 
\end{figure}

In this work, we study microscopic mechanism and pairing symmetry of superconductivity in multilayer graphene that develops from the spin- and valley-polarized quarter metal normal state. 
Our mechanism is based on the overscreening of Coulomb interaction due to charge fluctuations, which leads to an effective attraction at length scales on a fraction of the Fermi wavelength driving Cooper pairing.
Using a minimal model for the band dispersion, 
our theory predicts that superconductivity occurs at low densities with enhanced density of states due to the annular Fermi surface and displacement field-induced flatness of the band. 
Based on our analysis we further propose that a Wigner crystalline insulating state occurs at the Lifshitz transition from simply-connected to annular Fermi sea that splits the superconducting dome into two. 
This provides an explanation for the two proximate superconducting domes SC1 and SC2 with similar properties observed in the experiment \cite{Han2024Aug}.

In the spin- and valley-polarized state, the Pauli principle dictates that a Cooper pair can only be formed by two electrons having odd relative angular momentum, for example, with $p$- or $f$-wave symmetry \cite{Sigrist1991Apr}. 
Using Coulomb interaction and including the effect of dielectric screening in two dimensions, 
we find
robust $p$-wave superconductivity at densities and temperatures in reasonable agreement with the experiment. 
The calculated $T_c$ is on the order between $100\, {\rm mK}$ to a few ${\rm K}$, depending on the dielectric screening of the Coulomb interaction in the graphene film relative to the surrounding dielectric. 
Relatedly, we find that electrons are paired even relatively far from the Fermi surface.
Our results indicate that generally, a chiral 
$p - i \tau  p$
ordering is favored ($\tau = \pm 1$ stands for $K/K'$ valley, independent on spin polarization), and we predict a number of relevant experimental signatures for it.

\textit{Band dispersion.---}
In rhombohedral multilayer graphene, low-energy bands come from sublattice polarized states in the top and bottom layers. An out-of-plane electric field induces a potential bias equal to $2 D$ between these layers and opens up an energy gap while flattening the dispersion near $K$ and $K'$ points. Therefore, the Fermi energy is only a few ${\rm meV}$ above the band bottom for small electron density of around $5\times 10^{11} {\rm cm}^{-2}$, where time-reversal-breaking superconductivity is observed. 

The low-energy band dispersion is highly tunable by the electric field. As the displacement field $D$ increases, the curvature at $K$ and $K'$ changes from positive to negative \cite{Ghazaryan2021Dec,Ghazaryan2023Mar} as shown in Fig.~\ref{fig:1}(c) for ABCA tetralayer graphene, resulting in a Mexican-hat shaped dispersion. In this case, a Lifshitz transition from simply-connected to annular Fermi surface occurs 
when the Fermi energy crosses the top of the hat as the electron density is reduced.

We capture the essential features of the electric-field-tuned conduction band in 
rhombohedral $n$-layer graphene with a minimal band dispersion: 
\begin{equation}
    \varepsilon_k = D \sqrt{1 + (k/k_0)^{2n}} + \frac{\hbar^2 k^2}{2 m}
    \label{eq:dispersion}
\end{equation}
where we set $n=4$ corresponding to the tetralayer and treat $k_0$ and $m$ as  $D$-dependent fit parameters to approximate the 
band dispersion of multilayer graphene \footnote{We verified that qualitatively similar results for the superconducting order are obtained when the functional form of the dispersion is varied, as long as the main qualitative features are preserved.}, see App.~\ref{app:fits}.
The functional form of Eq.~\ref{eq:dispersion} is derived from an effective 2-band model 
of rhombohedral tetralayer graphene with nearest-neighbor hopping \cite{Koshino2009Oct,Slizovskiy2019Dec}. The dispersion \eqref{eq:dispersion} is circularly symmetric.
The inclusion of additional hopping terms leads to trigonal warping. 
For now, we neglect trigonal warping and Berry curvature effects, and will treat them perturbatively later.

\textit{Rytova-Keldysh potential.---} The density-density interaction can be written as 
\begin{eqnarray}
 H_{\rm int} = \frac{1}{2} \sum_{\bm k_1, \bm k_2, \bm q} V(\bm q) \psi^\dagger_{\bm k_1 + \bm q} \psi^\dagger_{\bm k_2 - \bm q} \psi_{\bm k_2} \psi_{\bm k_1} , 
 \label{eq:interaction}
\end{eqnarray}
where $\psi^{(\dagger)}_{k}$ are the annihilation (creation) operators of spin-polarized electrons in a single valley. 
Importantly, in a 2D material surrounded by a dielectric with a lower dielectric permittivity, the Coulomb interaction between two charges can be described by the Rytova-Keldysh potential~\cite{rytova2020screenedpotentialpointcharge,Keldysh1979Jun,Cudazzo2011Aug}, taking the form:
\begin{equation}
    V(\bm{q}) = \frac{e^2}{2 \epsilon|q|(1 + r_{K} |q|)} 
\end{equation}
where $\epsilon = 5 \, \epsilon_0$ 
is the dielectric permittivity of the surrounding hBN (with $\epsilon_0$ the vacuum permittivity), and $r_{K}$ is the Rytova-Keldysh parameter. 
The Rytova-Keldysh parameter depends on the difference between the dielectric response of the 2D material under study and that of the surrounding insulator.
Since the 2D dielectric screening is determined by interband transitions and thereby depends on the band gap, 
$r_{K}$ of multilayer graphene is affected by the displacement field \cite{Quintela2022Jul}. 
We will find that the superconducting pairing strength depends sensitively on $r_{K}$.

\begin{figure}[!t]
\includegraphics[width= 0.95\columnwidth]{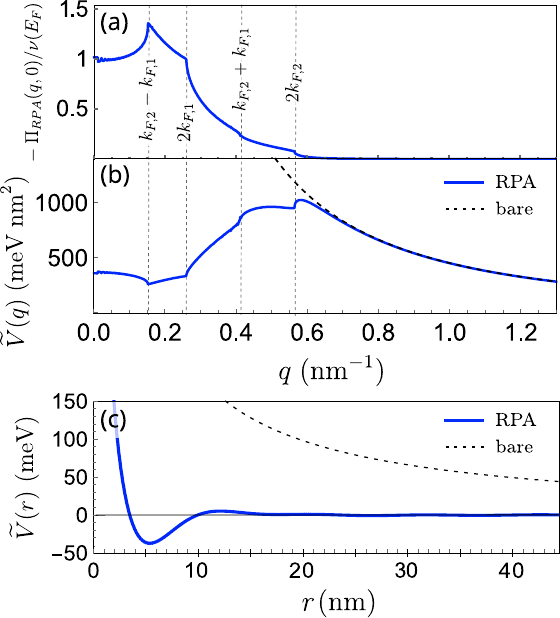}
\caption{(a) Charge susceptibility of the non-interacting electron gas for $D = 60 \, {\rm meV}$ at density $n = 0.5 \times 10^{12} \, {\rm cm}^{-2}$. The vertical lines indicate combinations of the momenta of the inner and outer Fermi surfaces $k_{F, 1}$ and $k_{F, 2}$. (b), (c) Screened (solid line) and bare (dashed) Rytova-Keldysh interaction potential in momentum and real space, respectively, with $r_K = 3\,{\rm nm}$.} 
\label{fig:2} 
\end{figure}

\textit{Electron pairing from screened Coulomb repulsion.---}
We study superconductivity in the spin- and valley-polarized state. This is justified because (i) valley polarization is observed to occur at a much higher temperature than superconductivity, and (ii)
the primary superconducting state (SC1 in Ref.~\cite{Han2024Aug}) was observed within the quarter-metal state, and (iii) the primary superconducting states (SC1 and SC2) survive to very high in-plane magnetic field. 

Theoretically, we also expect that the energy scales of spin- and valley polarization to be much larger than superconductivity: Estimating the energy scale of isospin polarization in the Stoner model \cite{Ghazaryan2023Mar} as $\Delta_{\rm pol} \approx \nu(E_F) V(2 k_{F,2}) E_F \approx 2 $ to $ 3\ {\rm meV}$ in the relevant range of density and dispacement field, it is an order of magnitude larger than the obtained pairing strength [Fig.~\ref{fig:4}]. 

Since total spin $S^2$ and valley imbalance $N_{K} - N_{K'}$ are conserved quantum numbers, there are no isospin fluctuations at zero temperature that can drive superconductivity in the fully-polarized quarter metal state. 
This motivates us to consider a mechanism for superconductivity in the spin- and valley-polarized state
 based on screening of electron-electron interactions by charge fluctuations \cite{Kohn1965Sep,Chubukov1993Jul,Maiti2013Aug}. The screening is described by the charge susceptibility $\chi_{e}^{{\rm R}}(\bm{q},\tau)=-\frac{1}{A}\langle T_{\tau}\rho_{e,\bm{q}}(\tau)\rho_{e,-\bm{q}}(0)\rangle$ which in random phase approximation (RPA) is determined by the expression
\begin{equation}
    \chi_{e}^{{\rm R}}(\bm{q},i\Omega_{n})\overset{\rm RPA}{\approx}\frac{\chi_{e}^{{\rm 0R}}(\bm{q},i\Omega_{n})}{1-V_{\bm{q}}\chi_{e}^{{\rm 0R}}(\bm{q},i\Omega_{n})/e^{2}}
\end{equation}
in terms of the charge susceptibility of the non-interacting electron gas
\begin{equation}
    \chi_{e}^{{\rm 0R}}(\bm{q},i\Omega_{n})=\frac{e^{2}}{A}\sum_{\bm{k}}\frac{n_{F}(\varepsilon_{\bm{k}})-n_{F}(\varepsilon_{\bm{k}+\bm{q}})}{\varepsilon_{\bm{k}}-\varepsilon_{\bm{k}+\bm{q}}+i\hbar\Omega_{n}}
    \label{eq:charge_susceptibility}
\end{equation}
where $\tau$ is imaginary time and $i\Omega_n$ Matsubara frequencies. 
With the resulting dielectric response function $\epsilon^{-1}(\bm{q},\omega)=1+V_{\bm{q}}\chi_{e}^{{\rm R}}(\bm{q},\omega)/e^{2}$ one obtains the screened interaction potential
\begin{equation}
    \tilde{V}_{\bm{q}} = V_{\bm{q}}\epsilon^{-1}(\bm{q},0) \overset{\rm RPA}{\approx} \frac{V_{\bm{q}}}{1-V_{\bm{q}}\chi_{e}^{{\rm 0R}}(\bm{q},0)/e^{2}}\,.
    \label{eq:interaction_RPA}
\end{equation}
The charge susceptibility and screened interaction potential for the annular Fermi pocket are shown in Fig.~\ref{fig:2}(a) and (b), respectively.
The $q \to 0$ limit of the denominator describing the screened interaction in Eq.~\eqref{eq:interaction_RPA}, 
$\lim _{q \to 0} V_{\bm q} \chi^{\rm 0R}_e(\bm q, 0) / e^2 = 2\pi / \ell_{\rm TF} q $ 
defines the Thomas-Fermi screening length $\ell_{\rm TF} = 4\pi\epsilon/e^2 \nu(E_F)$.

Due to the large density of states, the charge susceptibility becomes large for momenta below twice the Fermi momentum $k_{F,2}$ of the outer Fermi surface. This leads to a suppression of the electron-electron interaction at small momenta and a peak at $2 k_{F,2}$.
This renormalized interaction potential promotes a $p$-wave superconducting instability.

In real space, the screening strongly reduces the repulsion above the Thomas-Fermi screening length $\ell_{\rm TF}$. 
Furthermore, Friedel oscillations appear with period $2 k_{F,1},\  2 k_{F,2}$ due to the two Fermi surfaces. 
When the Thomas-Fermi screening length is much shorter than the wavelengths $\pi/k_{F,1},\  \pi/k_{F,2}$, Friedel oscillations dominate over Coulomb repulsion and lead to an effective attraction at short length scales, as obtained from our calculations shown in Fig.~\ref{fig:2}(c). We find $\ell_{\rm TF} \approx 1$ to $2\ {\rm nm}$, for typical parameters where we observe superconductivity as discussed below, while $2\pi/k_{F,1}$, $2\pi/k_{F,2}$ is typically around $20 \ {\rm nm} $ to $ 30\ {\rm nm}$. This mechanism benefits from a mexican-hat shaped dispersion because the additional pocket contributes a large density of states leading to a short Thomas-Fermi screening length. This enables superconductivity at larger density.

To determine the superconducting order parameter and its critical temperature, we solve the self-consistency equations
\begin{eqnarray}
    \Delta(\bm k) &=& \frac{1}{A}\sum_{\bm k'} \tilde{V}(\bm k - \bm k') \langle\hat{\psi}_{-\bm{k'}}\hat{\psi}_{\bm{k'}}\rangle, \nonumber \\
\langle\hat{\psi}_{-\bm{k}}\hat{\psi}_{\bm{k}}\rangle &=& -\frac{\Delta(\bm{k})}{2E_{\bm{k}}}\tanh\frac{E_{\bm k}}{2k_{B}T}, \label{eq:self-consistency-pairing}
\end{eqnarray}
where $E_{\bm k}$ is the quasiparticle energy 
\begin{eqnarray}
E_{\bm k} = \sqrt{(\varepsilon_{\bm k} -\mu)^2 + |\Delta(\bm k)|^2}.      
\end{eqnarray}
Here the interaction $\tilde{V}(\bm k - \bm k')$ scatters a pair of electrons at opposite momenta 
$(\bm k', -\bm k')$ to $(\bm k, -\bm k)$.    

\begin{figure}[!t]
\includegraphics[width= \columnwidth]{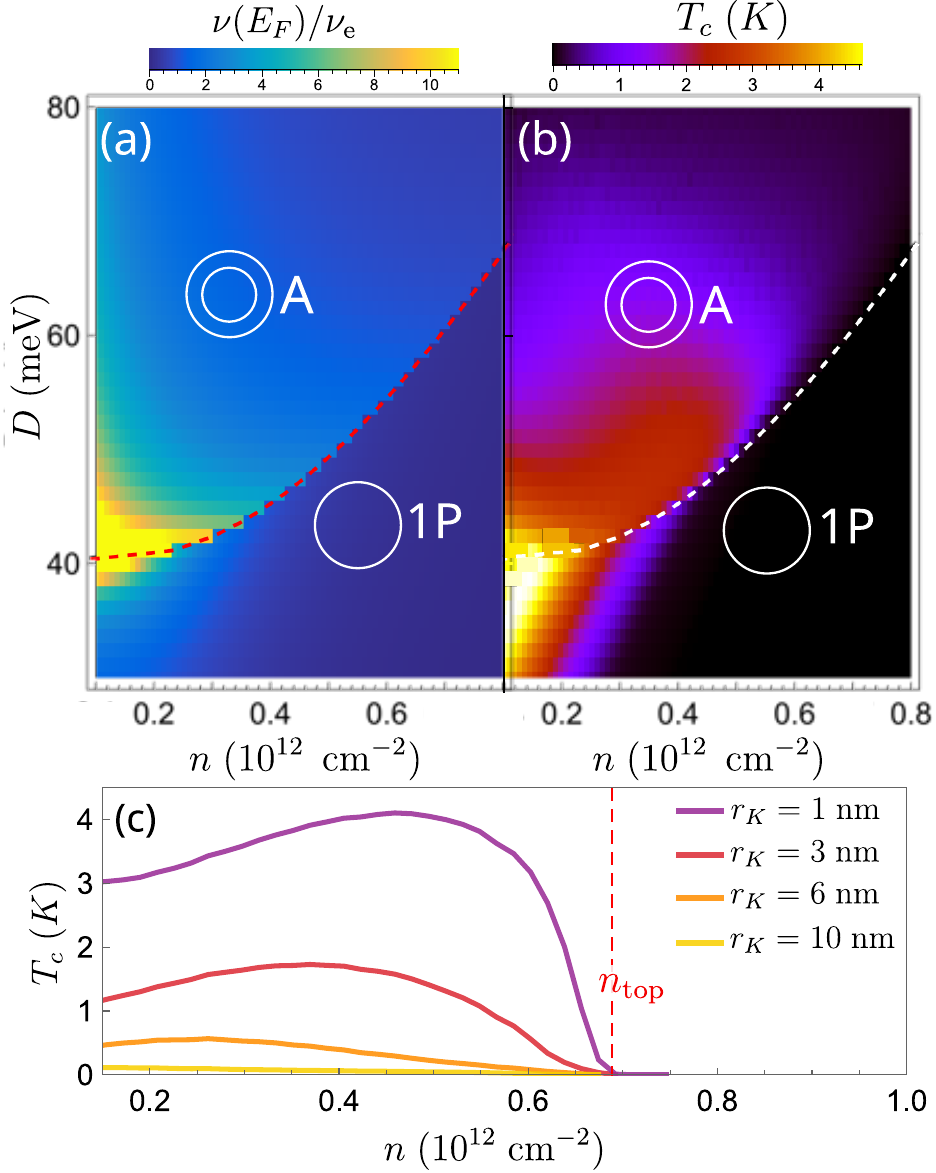}
\caption{ (a) Density of states of the minimal non-interacting model (Eq.~\eqref{eq:dispersion}) normalized by $\nu_{\rm e} = m_e /2\pi \hbar^2$ with $m_e$ the bare electron mass and (b) critical temperature of the chiral superconducting order as a function of electron density $n$ and displacement field $D$.  The dashed line indicates the Lifshitz transition from annular (A) to a single pocket (1P) Fermi sea. We choose $r_{K} = 3$ nm for the $T_c$ calculation. (c) Critical temperature for different dielectric screening parameterized by $r_{K}$ at $D = 60\,{\rm meV}$.}
\label{fig:3} 
\end{figure}

Due to the rotational symmetry in our model, we decompose the pairing interaction into angular harmonics: 
\begin{equation}
    \tilde{V}_{l}(k,k')\equiv \int_{0}^{2\pi}d\vartheta e^{il\vartheta}\tilde{V}(\vartheta,k,k')
    \label{eq:interaction-angular-harmonics}
\end{equation}
where we wrote $\tilde{V}_{\bm{k}-\bm{k}'}=\tilde{V}(\vartheta,k,k')$ with $\vartheta$ the angle between ${\bm k}$ and ${\bm k}'$ and $k,\, k'$ their magnitude. The order parameter can also be expanded into angular harmonics $
\Delta(\bm{k})=\sum_{l=0}^{\infty}\eta_{l}(k)e^{il\theta}
$
with $(k_x, k_y)=k(\cos \theta, \sin \theta)$. The equations for the critical temperature $T_c$ for different angular harmonics decouple. 

Around $T_c$, we linearize the self-consistency equation and find for the individual angular harmonics,
\begin{equation}
    \eta_{l}(k, T_{{\rm c}})=-\int_{0}^{\infty}\frac{dk'k'}{8 \pi^2}\tilde{V}_{l}(k,k')\frac{\tanh\frac{|\varepsilon_{k'} - \mu|}{2k_{B}T_{\rm c}}}{|\varepsilon_{k'}- \mu|}\eta_{l}(k', T_{{\rm c}})
    \label{eq:self-consistency-Tc}
\end{equation}
For the superconducting order parameter at zero temperature, the self-consistency relation reads
\begin{equation}
    \eta_{l}(k,0)	= -\int_{0}^{\infty}\frac{dk' k'}{8 \pi^2}\frac{\tilde{V}_{l}(k,k') \eta_{l}(k',0)}{\sqrt{(\varepsilon_{k'}-\mu)^{2}+\eta_{l}(k',0)^{2}}} .
    \label{eq:self-consistency-GapT0}
\end{equation}
Both equations can be solved iteratively, as detailed in App.~\ref{app:mean-field}
 \footnote{This angular decomposition applies for a circularly symmetric interaction potential, where the circular symmetry is preserved for the screened potential when the dispersion is circularly symmetric.}.

Importantly, the screened interaction $\tilde{V}_{\bm{q}}$ is positive and large at 
momentum transfer $\bm{q}=\bm{k}-\bm{k}'$ around $2k_{F,2}$, compared to small $\bm{q}$. 
Such $\bm{q}$-dependent interaction favors an order parameter that takes opposite signs at opposite $\pm \bm{k}$ points on the Fermi surface, i.e., it favors $p$-wave pairing.

For most superconductors, the pairing interaction is weak and therefore the pairing potential $\Delta(\bm k)$ is small compared to the Fermi energy and only appreciable in the vicinity of the Fermi surface. For this reason, in solving the gap equation, it suffices to use ``on-shell'' pairing interaction $\tilde{V}_l(k, k')$ at Fermi wavevectors  $k, k'$.  
In contrast, in multilayer graphene, the combination of low electron density and flat band bottom leads to a large ratio of interaction to the Fermi energy. As a consequence, we will show that $\Delta$ and $T_c$ can be on the order of the Fermi energy. Indeed, the recent experiment on tetralayer graphene~\cite{Han2024Aug} reports an unusually high upper critical field at at low density, indicating a strong-coupling superconductor with coherence length comparable to interparticle distance. For strong-coupling superconductors, the superconducting gap $\Delta(\bm k)$ can be large even away from Fermi momentum. Therefore, in solving the gap equation it is necessary to use the interaction $\tilde{V}_l(k, k')$ with full $k,k'$ dependence.

Our calculation of $T_c$ as a function of electron density and displacement field is shown in Fig.~\ref{fig:3}(b)]. 
Both the displacement field $D$ and the electron density $n$ together determine the Fermi surface size and topology.   
For small $D < D_c \approx 40\, {\rm meV}$, the dispersion increases monotonously with $k$ and the Fermi surface is a single circle with $k_F=\sqrt{4\pi n}$, whereas at large displacement field ($D > D_c$), a Lifshitz transition from annular to simply-connected Fermi sea occurs with increasing electron density, which is accompanied by a large jump in the normal-state density of states at the Fermi level [Fig.~\ref{fig:3}(a)]. 
Correspondingly, superconducting properties depend strongly on the displacement field.    
For small $D$, a superconducting state is found at small density where the chemical potential lies close to the relatively flat band bottom with large density of states. For large $D$,  
 the superconducting state sets in at low densities where the Fermi sea is annular, and $T_c$ is largest close to the Lifshitz transition.

It should be noted that the value of $T_c$ is sensitive to the dielectric screening of Coulomb repulsion by the multilayer graphene which, in turn, depends on the displacement field-induced band gap. As a function of the Rytova-Keldysh parameter $r_{K}$ [Fig.~\ref{fig:3}(c)], the typical $T_c$ at $D = 60\, {\rm meV}$ ranges from $T_c \approx 3\,{\rm K}$ at $r_{K} = 1\,{\rm nm}$ to a suppression of $T_c$ to below $0.1\,{\rm K}$ above $r_{K} = 10\,{\rm nm}$. 
With bare Coulomb repulsion, a maximal $T_c \approx 8\ {\rm K}$ is obtained, see App.~\ref{app:bare_coulomb}.
The decrease of $T_c$ with strong dielectric screening is consistent with electron pairing by Coulomb repulsion. Without knowing the value of $r_{K}$ for tetralayer graphene, we cannot make a quantitative prediction of $T_c$. 
Nonetheless, for the reasonable range of $r_{K}$ considered here, superconductivity always onsets at $n< 0.7 \times 10^{12}$ cm$^{-2}$ in agreement with experimental observation, and the calculated $T_c$ is acceptable compared to the experimental value, especially considering that the mean-field theory generally predicts higher values of $T_c$. 

Going beyond our mean-field treatment, we estimate the critical temperature for a BKT transition \cite{berezinskii1971destruction,Kosterlitz1973Apr,Halperin1979Sep} $k_B T_{\rm BKT} = \frac{\pi}{2} J$ determined by the superfluid stiffness. An order-of-magnitude estimate is obtained using the superfluid stiffness expression $J = \hbar^2 n/4m$ for a quadratic band whose mass $m \approx 2 m_e$ is matched so that its density of states matches the typical density of state $\nu(E_F) \approx 2 \nu_e$ at the Fermi energy of our model [compare Fig.~\ref{fig:3}(a)]. Based on this estimate, we find a typical BKT transition temperature $k_B T_{\rm BKT} \approx 0.9 \ {\rm K}$ at $n=0.5 \times 10^{12} \ {\rm cm}^{-2}$. This indicates that the critical temperature in this system may be limited by vortex proliferation in the BKT transition. 

\begin{figure}[!t]
\includegraphics[width= 1.00\columnwidth]{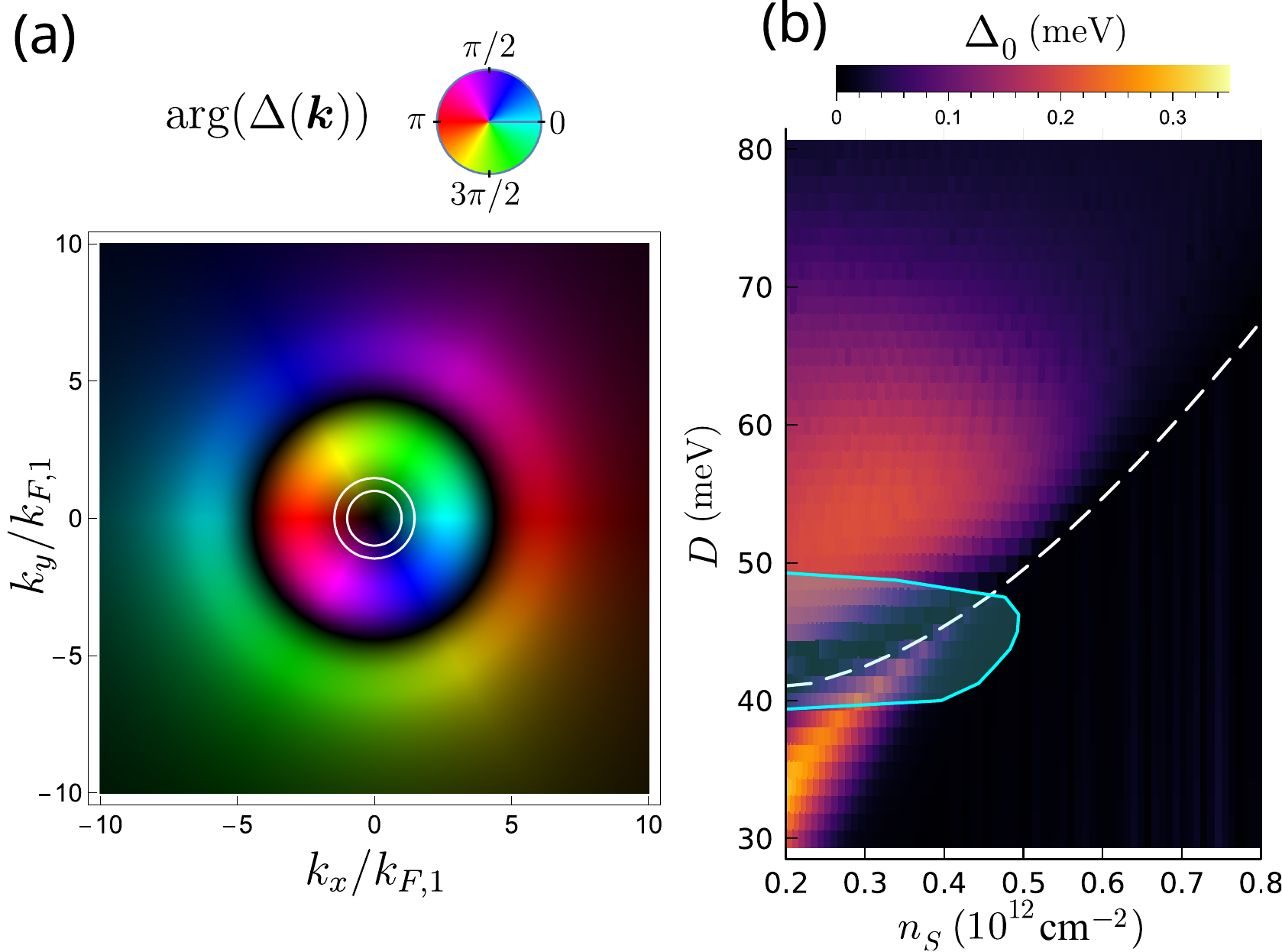}
\caption{(a) Momentum-dependent $p$-wave pairing potential, where the tone (from black to light) indicates its magnitude and the hue corresponds to the complex phase. The white circles show the Fermi surfaces. 
(b) Zero-temperature spectral gap $\Delta_0 = \min_k E_k$ as a function of displacement field and electron density $n_s$.  The dashed white line indicates the Lifshitz transition from annular (A) to a single pocket (1P) Fermi sea. The cyan line bounds a (shaded) region at small densities and moderate displacement fields where $r_s \geq 40$.}
\label{fig:4} 
\end{figure}

{\it Strong-coupling superconductivity.---}
For the chiral $p_x \pm ip_y$ pairing, the calculated pairing potential $\Delta(\bm{k})$ at $T=0$ can reach a few tenths of Fermi energy. Notably, $\Delta(\bm{k})$ is large not only close to the Fermi surface, but also extends to several times Fermi wavevector with a sign change at around $2k_{F,2}$ [Fig.~\ref{fig:4}(a)]. Its functional form closely follows the interaction potential up to a proportionality factor setting the pairing strength
[see App.~\ref{app:suppl-results}]. 
The presence of substantial pairing potential away from the Fermi surfaces is a consequence of the strong-coupling nature of this superconducting state -- because $k_B T_c$ and $\Delta$ are of order of the Fermi energy, pairing between electrons away from the Fermi surface is relevant. This is captured by our direct solution of Eqs.~\eqref{eq:self-consistency-Tc} and \eqref{eq:self-consistency-GapT0}. 

Due to the large pairing potential, the chemical potential changes appreciably in the superconducting state. 
This can be estimated by taking into account pairing potential in relating electron density to chemical potential
\begin{eqnarray}
 n_s = \int \frac{d{\bm k}}{4\pi^2} \left( 1 - \frac{\varepsilon_k - \mu}{E_k} \right).   
\end{eqnarray} 
At zero temperature, the change in chemical potential approaches $0.1 \, {\rm meV}$. The change in electron density is taken into account in Fig.~\ref{fig:4}(b). Near $T_c$, the pairing potential is strongly temperature dependent $\Delta\propto \sqrt{1-T/T_c}$, and therefore the change in chemical potential $\delta\mu \propto |\Delta|^2$ is expected to increase linearly with decreasing temperature.

{\it Topological superconductivity and Lifshitz transition.---} 
The quasiparticle gap $\Delta_0\equiv \min_k E_k$ as a function of density $n$ and displacement field $D$ is shown in Fig.~\ref{fig:4}(b). Our $p\pm i p$  superconductor generally has a full gap,  except at the Lifshitz transition from simply-connected to annular Fermi sea, where the system has a point node at the Fermi point $k = 0$ at which the  pairing potential vanishes. The closing of the quasiparticle gap marks a topological quantum phase transition from a topological superconductor with unity Chern number in the region with single Fermi pocket to a topologically trivial state in the annular region at large displacement fields \cite{Geier2020Jun,Kitaev2001Oct}. 
The topological superconductor at small displacement fields hosts chiral Majorana edge modes and Majorana zero modes in the vortex \cite{Read2000Apr}. In contrast, the trivial state with annular Fermi sea is adiabatically connected to the Bose-Einstein limit $\mu \to - \infty$, because the pairing potential $\Delta(\bm k)$ is finite at the band bottom of the Mexican hat dispersion which forms a ring at $k\neq 0$. Note that in our theory, even though the superconducting gap vanishes at $k=0$ at the Lifshitz transition, the gap at outer Fermi surface $\Delta(k_{F2})$ and critical temperature remains large throughout the transition-- therefore the quasiparticle gap closure is not visible in the critical temperature calculations Fig.~\ref{fig:3}(b).

\textit{Competing state.---}
Since the $p$-wave superconductivity found in our RPA calculation occurs at relatively low density, it is important to consider its competition with the Wigner crystal state, which generally appears in 2D Coulomb systems at sufficiently low density. 
To estimate the transition to Wigner crystal, we compute the gas parameter $r_s = E_{\rm int} / E_{\rm kin}$ given by the ratio of interaction $E_{\rm int}$ to kinetic energy $E_{\rm int}$ \cite{Drummond2009Mar}. In our calculations of $r_s$ here, we included a realistic distance $d = 30\, {\rm nm}$ to the metallic gates, whose screening modifies the bare interaction potential $V_{q} \to V_{q} \tanh q d$. We verified that this this screening does not visibly affect our calculations of $T_c$ and superconducting gap; however $r_s$ depends sensitively on the gate screening as the interparticle distances at low density approaches tens of nm.

In Fig.~\ref{fig:4}, the region where $r_s > 40$ is encircled by the cyan line; in a homogeneous electron gas the transition to a Wigner crystal occurs around $r_s \approx 30 - 40$ \cite{Tanatar1989Mar,Rapisarda1996,Drummond2009Mar,Spivak2004Oct,Monarkha2012Dec}. This region is where the band bottom is most flat, so that the kinetic energy per particle is low. 
It coincides with region where density of states is largest, see Fig.~\ref{fig:3}(a). 
We expect crystalline order is likely to dominate there, so that the superconducting region is divided in two, separated by an insulating Wigner crystal state.

\textit{Trigonal warping and Berry curvature effects.---}
Our minimal model neglects trigonal warping, which arises from electron hoppings beyond nearest neighbor atoms in multilayer graphene. With trigonal warping, the band dispersion becomes asymmetric $\epsilon_{\bm k} \neq \epsilon_{-\bm k}$, which weakens intravalley pairing between $\pm k$ states. Fortunately, the energy scale of trigonal warping is small in tetralayer graphene in the range of density and displacement field of interest, as evidenced by the nearly symmetric band dispersion shown in Fig.~\ref{fig:1}. Furthermore, our superconducting state driven by Coulomb interaction has a large gap up to a few  tenths of Fermi energy, and therefore is robust against the pair breaking effect of trigonal warping.      

Up to now we have neglected the effect of electron Bloch wavefunctions $\ket{u_{\bm k}}$ within the unit cell. The momentum dependence of complex-valued wavefunction $\ket{u_{\bm k}}$ gives rise to Berry curvature breaking time reversal symmetry, when the system is valley polarized. 
We now show that the Berry-phase effect generally favors a particular chirality for $p$-wave pairing within a given valley. 
To see this, we note that the {\it full} interaction term for the electrons in the conduction band is generally of the form
\begin{align}
    \tilde{H}_{\rm int} & = \frac{1}{2} \sum_{\bm k_1, \bm k_2, \bm q} V(\bm q) \bra{u_{\bm k_1+\bm q}} \ket{u_{\bm k_1}}\bra{u_{\bm k_2-\bm q}} \ket{u_{\bm k_2}} \nonumber \\
    & \quad \times 
    \widetilde{\psi}^\dagger_{\bm k_1 + \bm q} \widetilde{\psi}^\dagger_{\bm k_2 - \bm q} \widetilde{\psi}_{\bm k_2} \widetilde{\psi}_{\bm k_1}
\end{align}
where $\widetilde{\psi}^{(\dagger)}_{\bm k} $ creates an electron in the state $\ket{u_{\bm k}}$ in the conduction band.
Compared to Eq.~\eqref{eq:interaction}, the full interaction contains the form factor $\bra{u_{\bm k+\bm q}} \ket{u_{\bm k}}\bra{u_{\bm k'-\bm q}} \ket{u_{\bm k'}}$, which is complex-valued and thus breaks the time-reversal symmetry of the low-energy theory.

Anticipating zero-momentum Cooper pairing, we define $\bm{k}'=-\bm{k}_{1}-\bm{q}$ and focus on $ - \bm k_1 = \bm k_2 = \bm k$.
With this, we expand the form factor into harmonics
\begin{eqnarray}
 \bra{u_{-\bm k'}} \ket{u_{-\bm k}}\bra{u_{\bm k'}} \ket{u_{\bm k}} = \sum_l \alpha_l^\tau e^{-i l (\varphi_{\bm k'} - \varphi_{\bm k})}   
 \label{eq:form-factor}
\end{eqnarray}
where $\alpha_l^\tau = (\alpha_{-l}^{- \tau})^*$ due to time-reversal symmetry relating the two valleys $\tau = \pm 1$. 
The limit $\alpha_l \to 0$ for all $l \neq 0$ recovers our previous analysis using Eq.~\eqref{eq:interaction}. 
Now, we treat $\alpha_l$ with $l \neq 0$ as a perturbation to the $p$-wave superconducting state. 

Including the form factor, the condensation energy is
\begin{equation}
    E_{{\rm c}} =-\frac{1}{2A}\sum_{\bm{k},\bm{k}'}\tilde{V}_{\bm{k}-\bm{k}'}\bra{u_{-\bm k'}} \ket{u_{-\bm k}}\bra{u_{\bm k'}} \ket{u_{\bm k}}\langle\hat{\psi}_{-\bm{k}}\hat{\psi}_{\bm{k}}\rangle\langle\hat{\psi}_{\bm{k}'}^{\dagger}\hat{\psi}_{-\bm{k}'}^{\dagger}\rangle \, .
\end{equation}
For a pairing potential $\Delta(\bm k) = \eta_j(k) e^{ij\varphi_{\bm k}}$ with angular momentum $j$, 
using the angular decomposition of the form factor Eq.~\eqref{eq:form-factor} and the pairing interaction Eq.~\eqref{eq:interaction-angular-harmonics}, $E_{{\rm c}}$ can be expressed as 
\begin{equation}
    E_{{\rm c}} =-A \sum_{l}\alpha_{l}^{\tau}\int\frac{dkdk'}{2(2\pi)^{3}}kk'\tilde{V}_{-l-j}(k,k')\frac{\eta_{j}(k)\eta_{j}(k')}{4E_{\bm{k}}E_{\bm{k}'}}.
    \label{eq:condensation-energy-harmonics}
\end{equation}
When the form factor is a constant, only $\alpha_{0}$ term is present and $\tilde{V}_{j}=\tilde{V}_{-j}$ guarantees equal condensation energy for $\pm j$ pairings. However, with broken time reversal symmetry, $\alpha_{l\neq 0}$'s are generally nonzero and therefore the condensation energy is generally different for the two $p$-wave chiralities. In particular, a large contribution from the terms $j = -l$ proportional to $\tilde{V}_0(k,k')$ because $\tilde{V}_0(k,k')$ is positive definite. 

Within the two-band model of Ref.~\cite{Slizovskiy2019Dec} giving rise to dispersion Eq.~\eqref{eq:dispersion}, the wavefunction is of the form $|u(\bm k)\rangle = (1, \lambda e^{4 i \varphi_{\bm k}})$, so that $\alpha_4 \propto \lambda$ is the leading order correction in Eq.~\eqref{eq:form-factor}. Thus, in Eq.~\eqref{eq:condensation-energy-harmonics} for the condensation energy, the corrections for $j = \pm 1$ pairing are proportional to the angular harmonics $\tilde{V}_{3}$ and $\tilde{V}_5$, respectively, which lifts the degeneracy between the two chiralities. 
In App.~\ref{app:Berry_curvature_effects}, we calculate critical temperature for $l = \pm 1$ pairing including the $4\pi$ Berry phase from the model of Ref.~\cite{Slizovskiy2019Dec} without trigonal warping, and find a difference in critical temperature of around $20 \%$. 

\textit{Discussion.---}
We have shown that a strong-coupling chiral $p$-wave superconducting may emerge from charge fluctuations due to Coulomb repulsion in a spin- and valley-polarized state in multilayer graphene. 
The superconducting transition occurs at low density over a range of displacement fields where the band bottom is flat on the scale of the Fermi energy. 
In this range, increasing displacing field induces a Lifshitz transition from simply-connected to annular Fermi sea occurs which also marks a phase transition from a topological to a trivial superconducting state.
The chirality of the Bloch wave functions which is responsible for the Berry curvature selects the chirality of the $p$-wave superconducting order parameter. 

Our obtained critical temperature and density range is in rough agreement with a recent experiment in tetralayer graphene \cite{Han2024Aug}. 
In the experiment, quantum oscillations and anomalous Hall conductance measurements indicate the spin- and valley polarization. 
Superconductivity emerges in a region which does not show clear quantum oscillations, which indicates a large density of states (effective mass) in the relevant density range, consistent with our theoretical picture, see Fig.~\ref{fig:3}(a) and (b).

We also speculate that a charge ordered state may appear very close to the Lifshitz transition at low density, where the density of states and the ratio of interaction to kinetic energy are the largest. 
In this scenario, the ordered state divides the superconducting region into two domes. 
A similar feature has been observed in the experiment \cite{Han2024Aug}.

We expect that our mechanism may also apply to isospin-polarized phases of rhombohedral graphene with a different number of layers because all these systems share the feature of a large density of states at the band bottom under an applied displacement field \cite{Ghazaryan2023Mar}.
Crucially, calculations of Ref.~\cite{Ghazaryan2023Mar} indicate that tetralayer graphene reaches a significantly larger density of states close to the band bottom under a displacement field compared to smaller number of layers. 
Ref.~\cite{Ghazaryan2023Mar} also demonstrates that the isospin polarization emerges in a Stoner model 
due to large 
density of states and 
strong interaction.
Our theory shows that the large density of states induces a short Thomas-Fermi screening length that enables superconductivity from repulsion via overscreened Friedel oscillations. 
This significant quantitative difference may explain why superconductivity has been observed in rhombohedral tetralayer and pentalayer graphene \cite{Han2024Aug}, but not for smaller layer number up to now. 

Related, we remark that multilayer rhombohedral graphene under a displacement field exhibits a larger density of states close to the band bottom in the conduction compared to the valence band \cite{Ghazaryan2023Mar}, which favors the charge-fluctuation mechanism for superconductivity in the conduction but not in the valence band.

The RPA is expected to be qualitatively correct, but corrections beyond RPA are expected to significantly alter the quantitative predictions because of the absence of Migdal's theorem \cite{Migdal1958Jun} for superconductivity from screened electron-electron repulsion \cite{Takada1993Mar}. 
Different corrections contribute to an enhancement or reduction of $T_c$. 
A detailed study of corrections beyond RPA is left for future work.

Alternative mechanisms for superconductivity in an isospin-polarized metal include phonons and fluctuations of the isospin polarization. We assume that the superconductivity develops on top of a fully spin- and valley-polarized state, which is in line with experimental observation. For such case, the Pauli principle requires the superconducting state to be odd-parity. However, phonons typically lead to an isotropic attraction that favors the $s$-wave channel and is averaged out for higher angular momenta, and thus, we expect that this is likely not an important mechanism for pairing for the system considered here.  Fluctuations of the isospin order parameter can indeed mediate strong pairing in the proximity to the phase transition in the isospin order  \cite{Fay1980Oct,Lohneysen2007Aug,Dong2023May,Dong2023Oct}. 
We remark that these fluctuations are not present in the fully polarized phase that we consider here.

We note that, besides possible suppression of a charge-ordered competing state, screening from metallic gates becomes relevant to the superconducting state only when the distance to the metallic gates becomes of order of $\ell_{\rm TF}$. Because typical distances to metallic fields in experiment are of order $30\ {\rm nm}$, much larger than the typical $\ell_{\rm TF} \approx 1$ to $2 \ {\rm nm}$ obtained from our calculations, gate-screening is irrelavant to the superconducting state. Only when reducing the gate distance to the order of $\ell_{\rm TF}$ the superconducting order gets suppressed, see App.~\ref{app:gate_screening} for details. 

Finally, we note that the intravalley pairing implies a large Cooper pair momentum of $ \pm 2 \bm K$ 
which is commensurate with the lattice 
\cite{Fulde1964Aug,Larkin1964Sep,Roy2010Jul,Tsuchiya2016Sep,EunAhKim2017,Li2020Oct,Li2021Dec,Scammell2022May}. 
We also verified that our mechanism strongly favors $p$-wave pairing over higher angular momenta, see App.~\ref{app:f_h-wave}.

\textbf{Acknowledgements.}~
We thank Long Ju, Tonghang Han, Paco Guinea, Tommaso Cea, Erez Berg, Zhiyu Dong and Andrea Young for helpful discussions. This work was supported by a Simons Investigator Award from the Simons Foundation. M.G. acknowledges support from the German Research Foundation under the Walter Benjamin program (Grant Agreement No. 526129603). 
M.D. was supported in part by the Walter Burke Institute for Theoretical Physics at Caltech. L.F. was supported in part  by the U.S. Army DEVCOM ARL Army Research Office through the MIT Institute for Soldier Nanotechnologies under Cooperative Agreement number W911NF-23-2-0121.
The numerical calculations were performed using the Julia programming language \cite{Bezanson2017Feb}.

\textbf{Data availablility.}
The data to presented in this manuscript is available in the Zenodo repository {\tt Placeholder for link to zenodo repository}. 

\textbf{Code availablility.}
 The code used to compute the numerical results presented in this work is available in the GitHub repository \cite{spinpolarizedsuperconductivity_github}.

\textit{Note added:} During final stages of writing of the current manuscript, Ref.~\cite{Chou2024Sep} appeared, where a charge fluctuation mechanism for $p$-wave superconductivity in rhombohedral graphene has been discussed as well.

 \vfill

\bibliography{refs-KL.bib}

\clearpage 
\onecolumngrid

\begin{appendix}

\setcounter{figure}{0}
\renewcommand{\figurename}{Fig.}
\renewcommand{\thefigure}{S\arabic{figure}}

\section{Fit parameters for the dispersion}
\label{app:fits}

\begin{figure}
    \centering
    \includegraphics[width=0.8\linewidth]{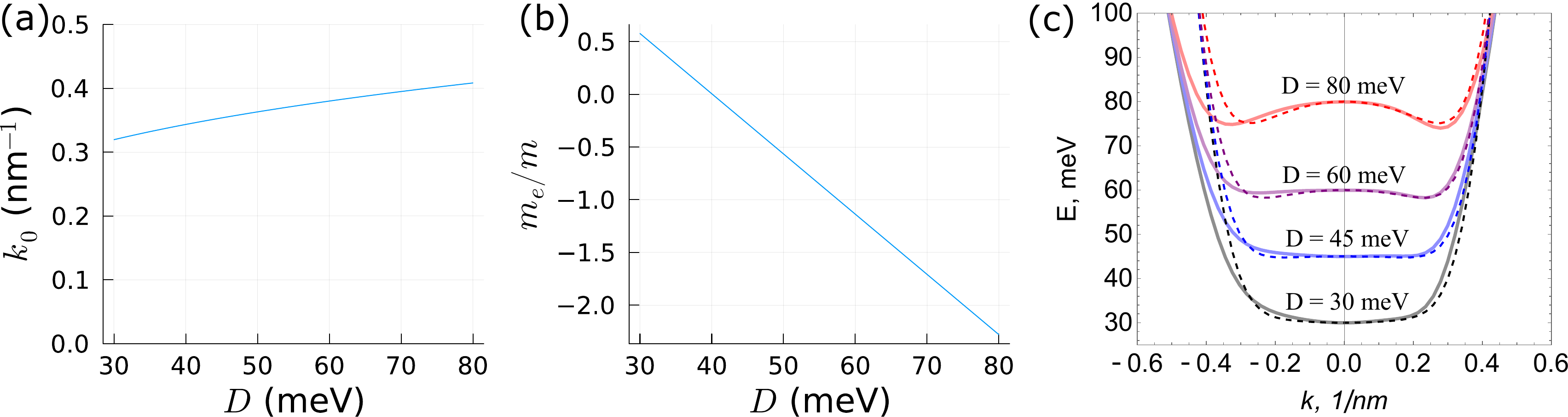}
    \caption{(a) Fit parameter $m$ and (b) $k_0$ of the Eq.~\eqref{eq:dispersion}. (c) Comparison of the minimal dispersion Eq.~\eqref{eq:dispersion} with fitted parameters (dashed lines) to the 8-band model from Ref.~\cite{Ghazaryan2021Dec,Ghazaryan2023Mar} (solid lines).}
    \label{fig:fits-dispersion}
\end{figure}

We fit our simplified dispersion Eq.~\eqref{eq:dispersion} to the dispersion of the realistic 8-band model from Refs.~\cite{Ghazaryan2021Dec,Ghazaryan2023Mar}. The obtained fit parameters as a function of $D$ and a comparison to the 8-band model are shown in Fig.~\ref{fig:fits-dispersion}. 

To validate our fit, we compare our density of states with the density of states obtained from the 8-band model in Ref.~\cite{Ghazaryan2023Mar}. 
Translating the result of Fig.~1(c) of Ref.~\cite{Ghazaryan2023Mar} to the spin- and valley polarized case (which requires to include a factor 1/4 in density of states and density), the density of states of the 8-band model with $D = 60\ {\rm meV}$ peaks at $n \approx 0.2\ 10^{12} {\rm cm}^{-2}$ with value $\nu \approx 5.5\ \nu_e$ and falls off to $\nu \approx 1.2\ \nu_e$ at $n \approx 0.5\ 10^{12} {\rm cm}^{-2}$. This behavior is approximately reproduced by the density of states of our circularly symmetric model, compare to Fig.~\ref{fig:3}(a) in the main text. 

\section{Mean-field theory}
\label{app:mean-field}

We perform the usual mean-field treatment and obtain the self-consistency equation:
\begin{equation}
    \Delta(\bm k) = \frac{1}{A}\sum_{\bm k'} V(\bm k - \bm k') \langle \psi_{-\bm k'} \psi_{\bm k'}\rangle
\end{equation}
where for a ${\rm SO}(2)$ rotation-symmetric dispersion 
\begin{equation}
\langle\hat{\psi}_{-\bm{k}}\hat{\psi}_{\bm{k}}\rangle=-\frac{\Delta(\bm{k})}{2E_{\bm{k}}}\tanh\frac{E_{\bm{k}}}{2k_{B}T}
\end{equation}
%
where $E_{\bm k} = \sqrt{(\varepsilon(\bm k) - \mu)^2 + |\Delta_{\bm k}|^2}$.

Due to the ${\rm SO}(2)$ rotation symmetry, the pairing potential can be expanded in angular harmonics 
\begin{equation}
    \Delta(\bm{k})=\sum_{m=0}^{\infty}\eta_{m}(k)e^{im\theta}
    \label{eq:Delta-angular-decomposition}
\end{equation}
where the equations for the critical temperature $T_c$ for different angular harmonics decouple, as we show below.

\subsubsection{Linearized gap equation for solutions around $T_{\rm c}$}

We describe a numerical procedure to obtain $T_{{\rm c},m}$ and the radial profile of the pairing potential $\eta_m(k)$ for different angular harmonics $m$.

Searching for solutions around the critical temperature $T_{\rm c}$, we may linearize in $\Delta({\bm k})$ so that $E_{\bm{k}}\approx|\xi_{\bm{k}}-\mu|$ and 
\begin{equation}
    \Delta(\bm{k})=-\frac{1}{A}\sum_{\bm{k}'}\tilde{V}_{\bm{k}-\bm{k}'}\frac{\Delta(\bm{k}')}{2|\xi_{k'}-\mu|}\tanh\frac{|\xi_{k'}-\mu|}{2k_{B}T}
\end{equation}
Expanding the pairing potential into angular harmonics, Fourier transformation reveals that different angular harmonics decouple, and we obtain Eq.~\eqref{eq:self-consistency-Tc} from the main text.

Eq.~\eqref{eq:self-consistency-Tc} equation can be solved iteratively as follows. Starting from an initial guess $\eta_{m}^{(0)}(k)$ (and for any iteration $\eta_{m}^{(n)}(k)$) for which we require only normalization $\int_{0}^{\infty}\eta_{m}^{(n)}(k)dk=1$, we obtain a corresponding $T_{{\rm c},m}^{(n)}$ by integrating over $\int_{0}^{\infty}dk_{1}$ to obtain a closed form expression,
\begin{equation}
    1=-\frac{1}{2(2\pi)^{2}}\int_{0}^{\infty}dk\int_{0}^{\infty}dk'k'\tilde{V}_{m}(k,k')\eta_{m}^{(n)}(k') \frac{\tanh\frac{|\xi_{k'}|}{2k_{B}T_{{\rm c},m}^{(n)}}}{|\xi_{k'} - \mu|}
\end{equation}
where for the second line we used the symmetry $\tilde{V}_{m}$. Having found $T_{{\rm c},m}^{(n)}$, we obtain the next iteration 
\begin{equation}
    \eta_{m}^{(n+1)}(k)=-\frac{1}{2(2\pi)^{2}}\int_{0}^{\infty}dk'\tilde{V}_{m}(k,k')\frac{\tanh\frac{|\xi_{k'}|}{2k_{B}T_{{\rm c},m}^{(n)}}}{|\xi_{k'} - \mu|}\eta_{m}^{(n)}(k')
\end{equation}
for which $T_{{\rm c},m}^{(n+1)}$ can then again be found using the above implicit equation. We found that the functional form of $\eta_{m}(k)$ is approximated by $\tilde{V}_m(k_{F,2},k)$ up to a factor setting the overall magnitude [compare to Fig.~\ref{fig:supp-gaps}(a)], so that $\eta_{m}^{(0)}(k) = \tilde{V}_m(k_{F,2},k)$ is a good initial guess.

\subsubsection{Self-consistency for zero-temperature gap}

At zero temperature, the self-consistency equation simplifies
\begin{equation}
    \Delta(\bm{k})=-\frac{1}{2A}\sum_{\bm{k}'}V_{\bm{k}-\bm{k}'}\frac{\Delta(\bm{k}')}{E_{\bm{k}'}}
\end{equation}
Formally, this equation does not separate into individual angular harmonics $\eta_{m}(k)$ of the scattering potential. However, we anticipate that decoupled solutions solve the self-consistency equation and solve for the zero-temperature pairing potential of each angular momentum separately. This directly leads to Eq.~\eqref{eq:self-consistency-GapT0} in the main text.

Eq.~\eqref{eq:self-consistency-GapT0} equation can again be solved iteratively. With the Ansatz  $\eta_{m}^{(0)}(k)=|\eta_{m}^{(0)}|f_{m}^{(0)}(k)$ in terms of  magnitude $|\eta_{m}^{(0)}|$ and functional form $f_{m}^{(0)}(k)$ for which we require normalization $\int_{0}^{\infty}f_{m}^{(0)}(k)dk = 1$, the magnitude is found self consistently by integration over $k$,
\begin{equation}
    1=-\frac{1}{2(2\pi)^{2}}\int dk\int_{0}^{\infty}dk'k'V_{m}(k,k')\frac{f_{m}^{(0)}(k')}{||f_{m}^{(0)}||}\frac{1}{\sqrt{(\xi_{k'}-\mu)^{2}+\left(|\eta_{m}^{(0)}|f_{m}^{(0)}(k')\right)^{2}}}\, .
\end{equation}
The functional form of the next iteration is then defined as
\begin{equation}
    f_{m}^{(n+1)}(k)=- {\cal N}_m^{(n+1)}\frac{1}{2(2\pi)^{2}}\int_{0}^{\infty}dk'k'V_{m}(k,k')\frac{f_{m}^{(n)}(k')}{\sqrt{(\xi_{k'}-\mu)^{2}+\eta_{m}^{(n)}(k')^{2}}}
\end{equation}
up to normalization, for which one then again determines the magnitude $|\eta_{m}^{(n)}|$ as above. A good ansatz for the starting point $f_{m}^{(0)}$ is the functional form $\eta_m(k)$ obtained at the critical temperature, or in terms of $\tilde{V}_m(k_{F,2},k)$ [see above and Fig.~\ref{fig:supp-gaps}(a)].

\section{Zero-temperature gap}
\label{app:suppl-results}

\begin{figure}
    \centering
    \includegraphics[width=0.5\linewidth]{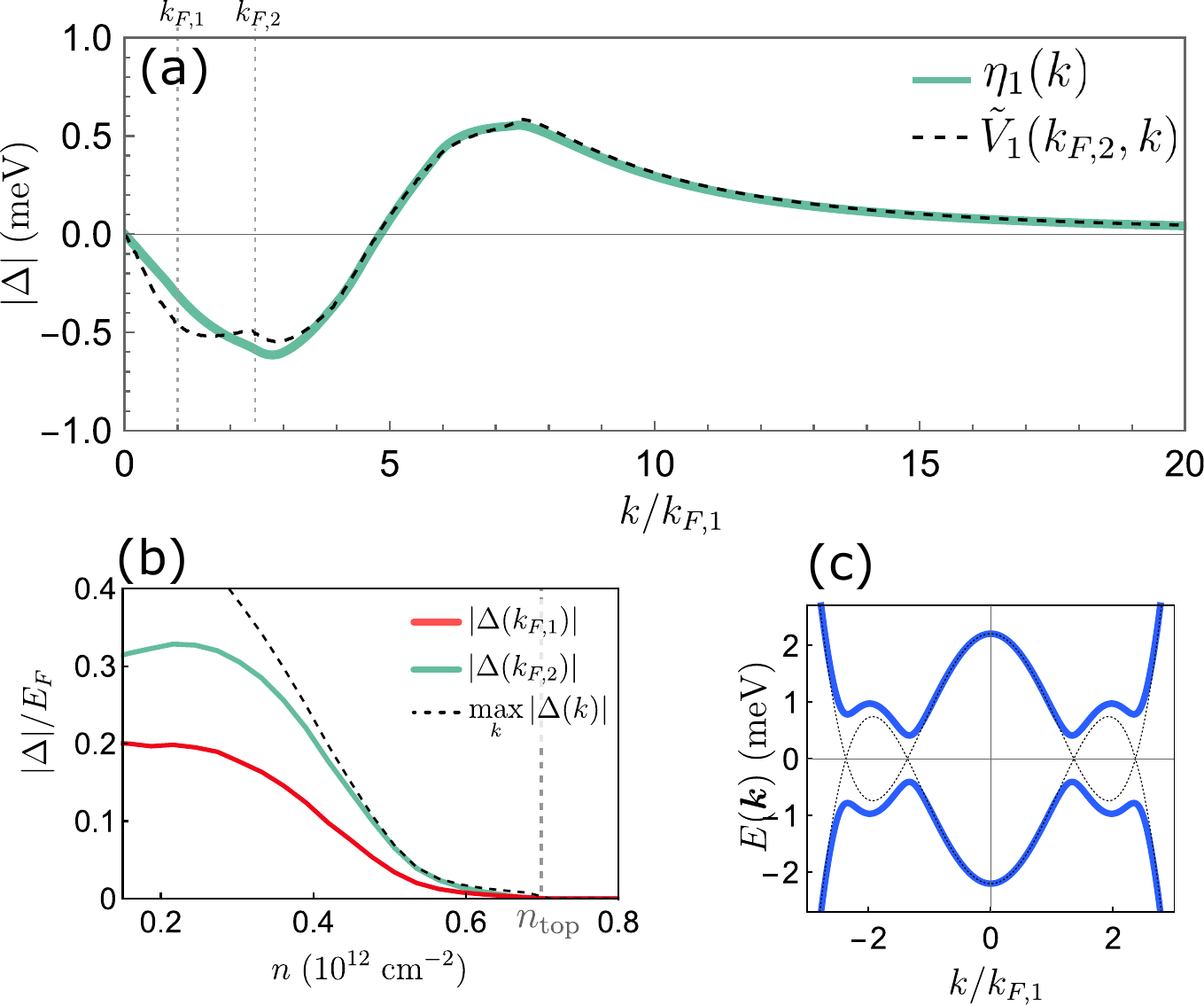}
    \caption{(a) Pairing potential $\eta_1(k)$ as a function of radial momentum $k$ and first harmonic of the interaction $\tilde{V}_1(k_{F,2}, k)$ for density $n = 0.5\, 10^{12} \, {\rm cm}^{-2}$. (b) Pairing potential at the inner ($k_{F,1}$) and outer ($k_{F,2}$) Fermi surface and maximal magnitude $\max_k |\eta_1(k)|$. (c) Zero-temperature dispersion of Bogoliubov quasiparticles at $n = 0.5\, 10^{12} \, {\rm cm}^{-2}$. Here we use the displacement field $D = 60 \, {\rm meV}$ and Keldysh parameter $r_{K} = 3\, {\rm nm}$.}
    \label{fig:supp-gaps}
\end{figure}

A line plot of the radial profile of the zero-temperature superconducting gap is shown in Fig.~\ref{fig:supp-gaps}(a). Interestingly, the radial profile of the superconducting gap follows closely the functional form $V_{l = 1}(k_{F,2},k)$ of the first harmonic of the scattering potential when one of the scattering partners has momentum $k_{F,2}$ at the outer Fermi surface. 
Additionally, the radial profile of the superconducting gap at $T_c$ is very similar to the functional form at zero temperature (data not shown).

Fig.~\ref{fig:supp-gaps}(b) shows the pairing potential at the inner and outer Fermi surface, as well as its maximum as a function of density, at the same parameters as the critical temperature data in Fig.~\ref{fig:3}(c) at $r_{K} = 3\, {\rm nm}$. The ratio $\eta_1(k_{F,2} / k_B T_c$ slightly depends on density but remains close to two. The Bogoliubov quasiparticle dispersion at zero temperature and density $n = 0.5\, 10^{12} \, {\rm cm}^{-2}$ is shown in Fig.~\ref{fig:supp-gaps}(c). 

\section{Gas parameter $r_s$}
\label{app:r_s}

\begin{figure}
    \centering
    \begin{tabular}{cc}
         (a) $ d = 30\, {\rm nm}$ & (b) $d \to \infty $ \\
         \includegraphics[width=0.35\linewidth]{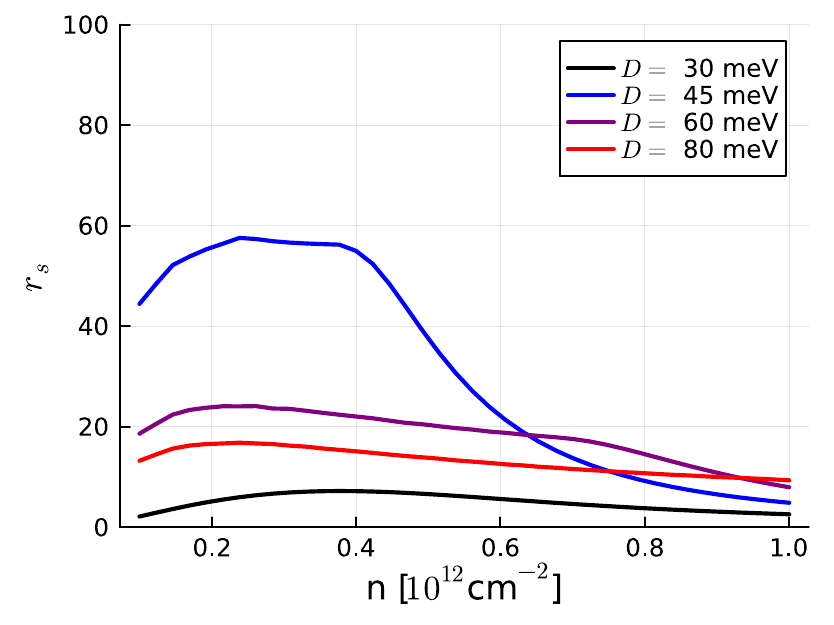} & \includegraphics[width=0.35\linewidth]{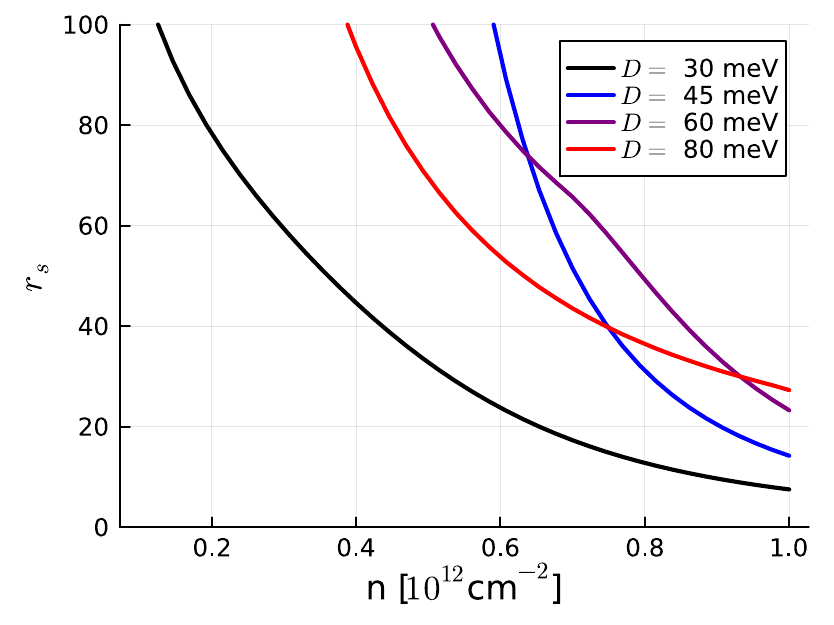}
    \end{tabular}
    \caption{Gas parameter $r_s$ for distance to the metallic gates of (a) $d = 30\, {\rm nm}$ and (b) $d \to \infty$. }
    \label{fig:supp-r_s}
\end{figure}

Fig.~\ref{fig:supp-r_s} (a) shows the gas parameter $r_s$ as a function of density and displacement field for a distance of $d = 30\, {\rm nm}$ to the gates. Only when the dispersion is flat, the gas parameter reaches large values above 30, where a transition to a charge-ordered state typically occurs. 
The gas parameter is significantly suppressed by screening from the gates -- when the gates are taken infinitely far away, the gas parameter is much larger, compare to Fig.~\ref{fig:supp-r_s} (b). This suggests that screening from the metallic gates plays an important role in suppressing a competing charge-ordered state. 

\section{Berry curvature effects on superconductivity}
\label{app:Berry_curvature_effects}

The Berry curvature originates from the projection of the interaction onto the conduction band. The quantum geometry associated to the projection enters the calculation of the superconducting properties on two stages.

First, the charge susceptibility of the non-interacting Fermi gas in the conduction band is modified,
\begin{equation}
    \chi_{e}^{{\rm 0R}}(\bm{q},i\Omega_{n})=\frac{e^{2}}{A}\sum_{\bm{k}}|\langle u_{\bm k} |u_{\bm k + \bm q}\rangle|^2\frac{n_{F}(\varepsilon_{\bm{k}})-n_{F}(\varepsilon_{\bm{k}+\bm{q}})}{\varepsilon_{\bm{k}}-\varepsilon_{\bm{k}+\bm{q}}+i\hbar\Omega_{n}} \label{eq:charge_susceptibility_form-factor}
\end{equation}
This equation replaces Eq.~\eqref{eq:charge_susceptibility} in the main text. This modified charge susceptibility enters the RPA calculation of the renormalized interaction, Eq.~\eqref{eq:interaction_RPA}.

Second, the self-consistency relation for the superconducting pairing acquires the form
\begin{eqnarray}
    \Delta(\bm{k})=\frac{1}{A}\sum_{\bm{k}'}\tilde{V}_{\bm{k}-\bm{k}'}\langle u_{\bm{k}}|u_{\bm{k}'}\rangle\langle u_{-\bm{k}}|u_{-\bm{k}'}\rangle\frac{\Delta(\bm{k}')}{2E_{\bm{k}'}}\tanh\frac{E_{\bm{k}'}}{2k_{B}T}
\end{eqnarray}
This equation replaces Eq.~\eqref{eq:self-consistency-pairing} in the main text. The form factors can be included in the angular decomposition of the renormalized interaction,
\begin{equation}
    \tilde{V}_{l}(k,k')\equiv \int_{0}^{2\pi}d\vartheta e^{il\vartheta}\langle u_{\bm{k}}|u_{\bm{k}'}\rangle\langle u_{-\bm{k}}|u_{-\bm{k}'}\rangle\tilde{V}(\vartheta,k,k')
    \label{eq:interaction-angular-harmonics_form-factor}
\end{equation}
This equation replaces Eq.~\eqref{eq:interaction-angular-harmonics} in the main text. 
The self-consistency equations Eq.~\eqref{eq:self-consistency-Tc} and \eqref{eq:self-consistency-GapT0} then take the same functional form, but with the modified angular harmonics $\tilde{V}_{l}(k,k')$ of the renormalized interaction defined in Eq.~\eqref{eq:interaction-angular-harmonics_form-factor}.
Because the form factors are complex, the symmetry between positive and negative angular harmonics $\pm l$ of the interaction is broken. This implies the selection of one of the angular momenta of the superconducting pairing over the other. 

In our system, the electronic quasiparticles in the $K$ and $K'$ valleys have an orbital magnetic moment associated to the Berry curvature in the valleys \cite{Xiao2010Jul}. 
Neglecting trigonal warping, a low-energy around the $K$ and $K'$ points is given by \cite{Koshino2009Oct,Slizovskiy2019Dec}
\begin{equation}
    H_{{\rm ABCA}}=\left(\begin{array}{cc}
D+\frac{\hbar^{2}k^{2}}{2m} & D\frac{k^{4}}{k_{0}^{4}}e^{\xi 4 i\gamma}\\
D\frac{k^{4}}{k_{0}^{4}}e^{- \xi 4i\gamma} & -D+\frac{\hbar^{2}k^{2}}{2m}
\end{array}\right)
\label{eq:eff_H_without_trig}
\end{equation}
where $(k_x, k_y) = (k \cos \gamma, k\sin \gamma)$ and $\xi = 1$ ($-1$) indicates the $K$ ($K'$) valley. 
The spectrum of the condition band of this Hamiltonian equals our effective model, Eq.~\eqref{eq:dispersion}. The conduction band eigenstate and form factor is
\begin{align}
    |u_{\bm k}\rangle&=\left(\begin{array}{c}
\cos\frac{\theta}{2}\\
e^{-\xi 4i\gamma}\sin\frac{\theta}{2}
\end{array}\right) \\
\langle u_{\bm{k}'}|u_{\bm{k}}\rangle&=\cos\frac{\theta'}{2}\cos\frac{\theta}{2}+e^{\xi i4\gamma' -\xi i4\gamma}\sin\frac{\theta'}{2}\sin\frac{\theta}{2}
\end{align}
where $\cos\theta=(1+k^{8}/k_0^8)^{-1/2}$. 

\begin{figure}
    \centering
    \begin{tabular}{ll}
         (a) \quad $D = 60 \ {\rm meV}$ & (b) \quad $D = 60 \ {\rm meV}$ \\
         \includegraphics[width=0.35\linewidth]{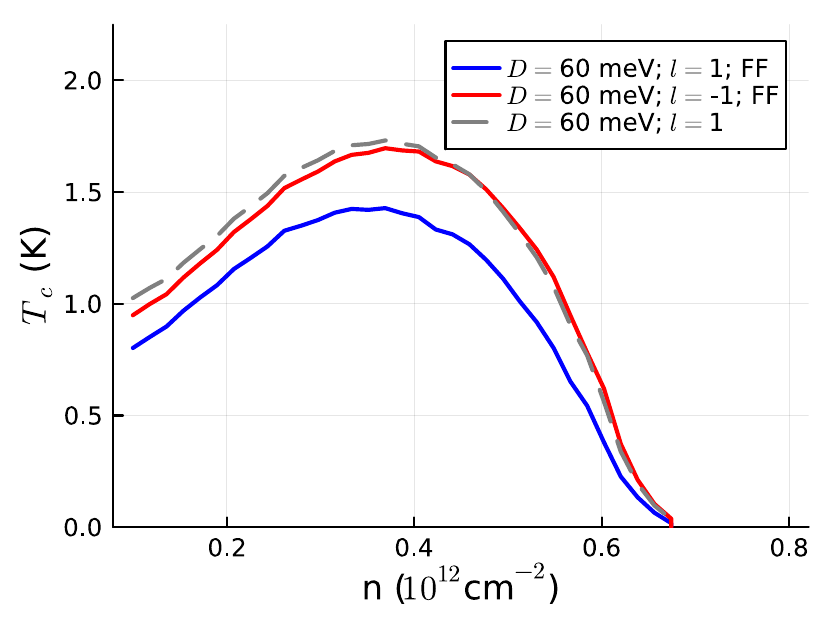} & \includegraphics[width=0.35\linewidth]{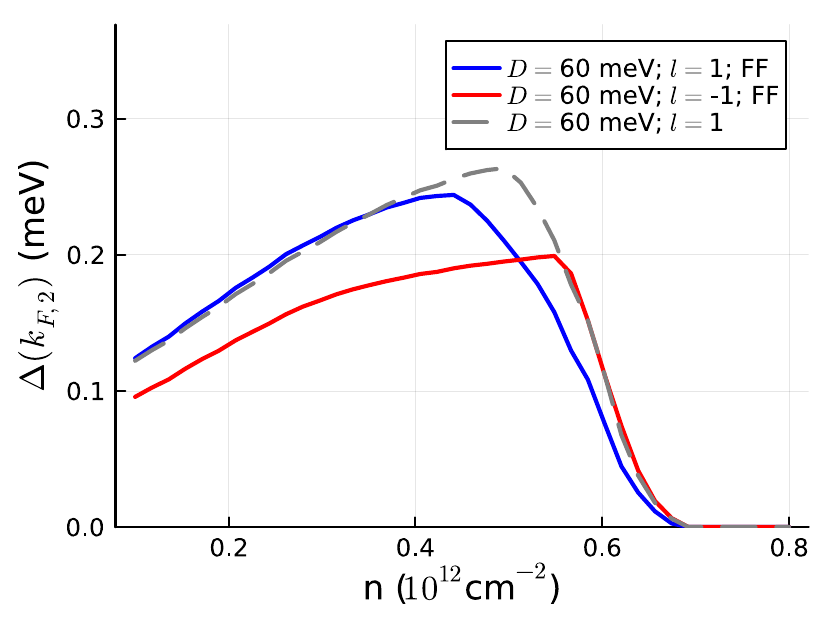} \\
         (c) \quad $D = 35 \ {\rm meV}$ & (d) \quad $D = 35 \ {\rm meV}$ \\
         \includegraphics[width=0.35\linewidth]{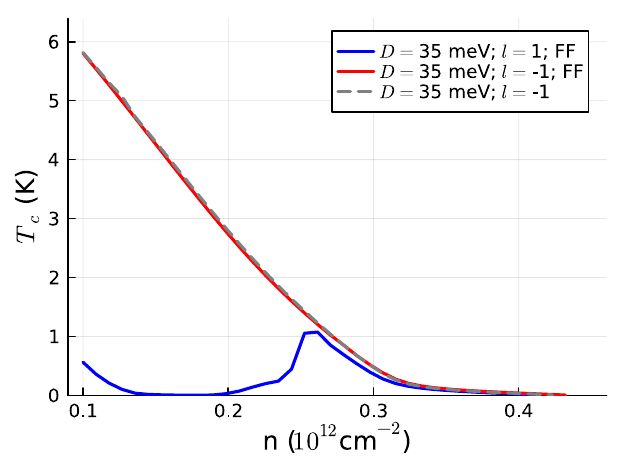} & \includegraphics[width=0.35\linewidth]{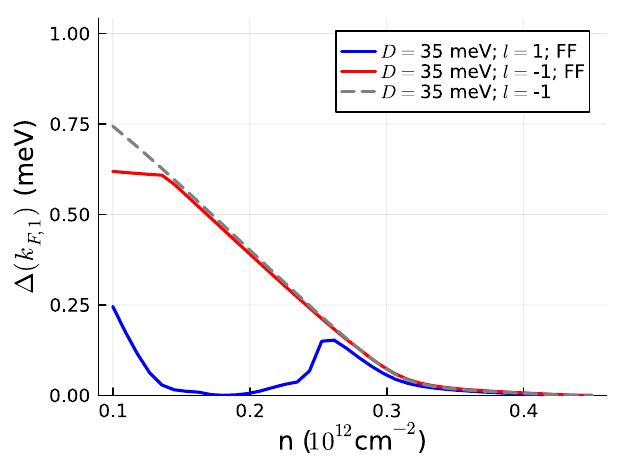}
    \end{tabular}
    \caption{Critical temperature [(a), (c)] and zero-temperature pairing potential at the outer Fermi surface $k_{F,2}$ [(b), (d)] at $D = 60\, {\rm meV}$  [(a), (b)] and $D = 35\, {\rm meV}$  [(c), (d)] for $l = 1$ (blue) and $ -1$ (red) angular momentum pairing including the form factor (FF) from the $K$ valley ($\xi = 1$) at $D = 60\ {\rm meV}$. The result without form factor (gray dashed) is shown for comparison. For $D = 60\, {\rm meV}$  [(a), (b)] the system is in the topologically trivial chiral superconducting emerging in the system with annular Fermi sea, while for $D = 35\, {\rm meV}$  [(a), (b)] the system is in the topological chiral superconducting phase with fully connected Fermi sea. Here we used a gate screening distance of $d = 30 \ {\rm nm}$.}
    \label{fig:supp-berry-phase}
\end{figure}

Fig.~\ref{fig:supp-berry-phase} shows the calculated critical temperature and zero-temperature pairing potential at the outer Fermi surface including the form factor. 

For $D = 60\, {\rm meV}$ [Fig.~\ref{fig:supp-berry-phase} (a), (b)], i.e. when the Fermi sea is annular and the chiral superconducting state is topologically trivial, we find that the Berry phase effects arising from the $4 \pi $ winding of the effective theory Eq.~\eqref{eq:eff_H_without_trig} without trigonal warping lead to an imbalance of around $\approx 20\%$ in the critical temperature between $l = \pm 1$ pairing. We expect that a detailed treatment including trigonal warping may lead to a larger difference between $l = \pm 1$ pairing, because trigonal warping induces terms with $\pi$ Berry phase winding around the Dirac cone. These terms perfectly cancel with the opposite angular momentum in the angular momentum decomposition of the interaction, Eq.~\eqref{eq:interaction-angular-harmonics_form-factor}, yielding a large contribution. 

For $D = 35\, {\rm meV}$ [Fig.~\ref{fig:supp-berry-phase} (c), (d)], i.e. when the Fermi sea is simply connected and the chiral superconducting state is topological, we find that both $T_c$ and pairing strength of one of the two pairing chiralities is strongly suppressed. This demonstrates that the orbital angular momentum of the electronic quasiparticles in the valleys strongly favors one pairing over the other.

\section{Dependence on screening by metallic gates}
\label{app:gate_screening}

When the graphene film is enclosed between metallic gates on both sides, the static Coulomb repulsion is modified
\begin{eqnarray}
    V(\bm q) \to V(\bm q) \tanh q d
\end{eqnarray}
where $d$ is the distance from the film to the metallic gates, assumed equal on both sides. We note that this approximation holds only in the static limit, {\it i.e.} when the Fermi velocity of the charge carriers in the metallic gates is much larger than the Fermi velocity of charges in the film. 

\begin{figure}
    \centering
    \begin{tabular}{ll}
         (a)  & (b) \\
         \includegraphics[width=0.35\linewidth]{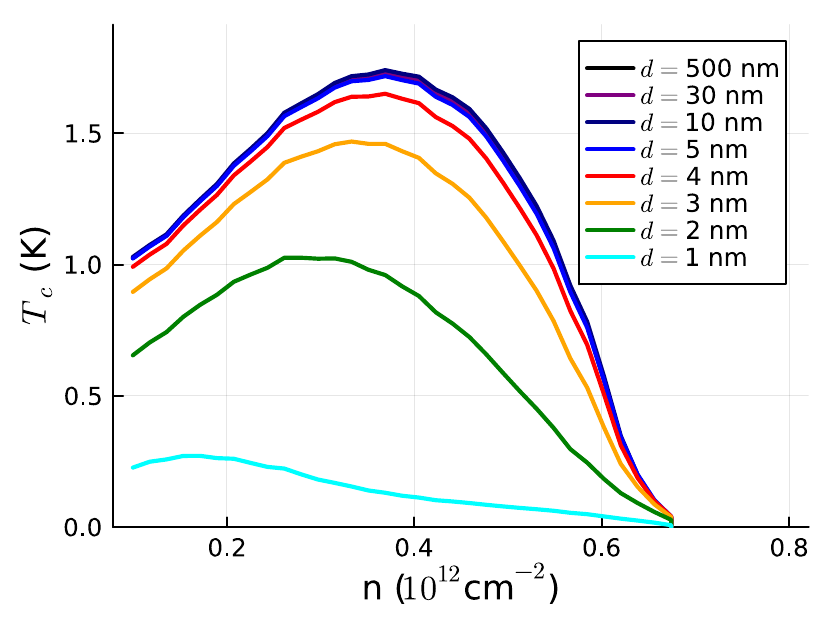} & \includegraphics[width=0.35\linewidth]{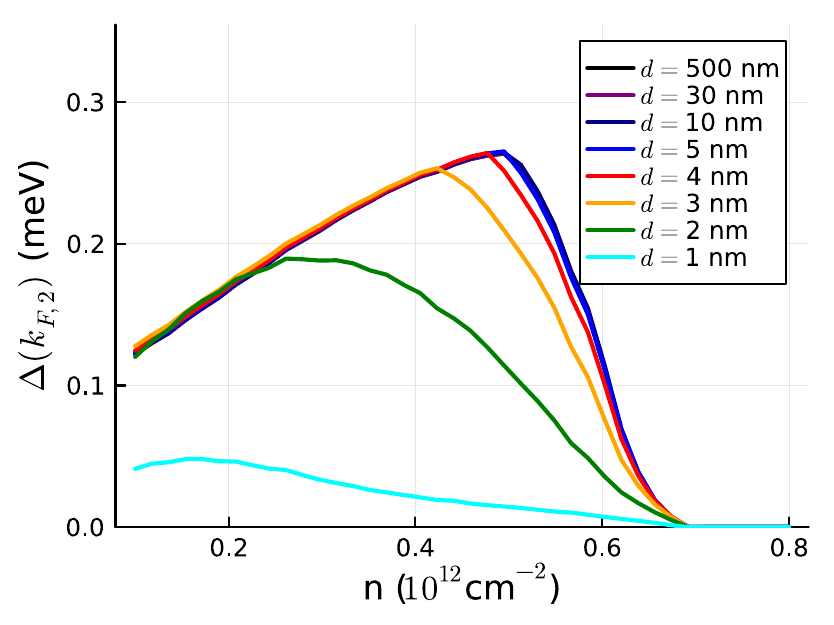}
    \end{tabular}
    \caption{Critical temperature (a) and zero-temperature pairing potential at the outer Fermi surface $k_{F,2}$ (b) at $D = 60\, {\rm meV}$ as a function of distance $d$ to the metallic gates. This calculation does not include form factors.}
    \label{fig:supp-d_gates}
\end{figure}

Our results [Fig.~\ref{fig:supp-d_gates}] show that the distance to the gates becomes relevant only on scales below the dynamic screening length $\approx 4\ {\rm nm}$ in our case. When the metallic gates are closer than this length scale, they suppress the attraction arising from the Friedel oscillations of the dynamically screened interaction in the 2DEG. 

\section{Higher angular momentum pairing}
\label{app:f_h-wave}

\begin{figure}
    \centering
    \begin{tabular}{cc}
         (a) $D = 30 \, {\rm meV}$ & (b) $D = 60 \, {\rm meV}$ \\
         \includegraphics[width=0.35\linewidth]{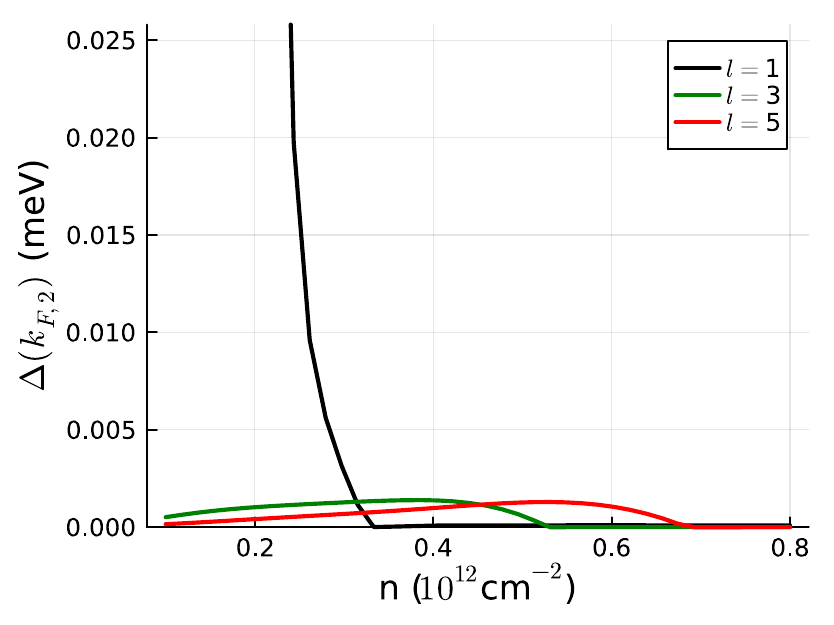} & \includegraphics[width=0.35\linewidth]{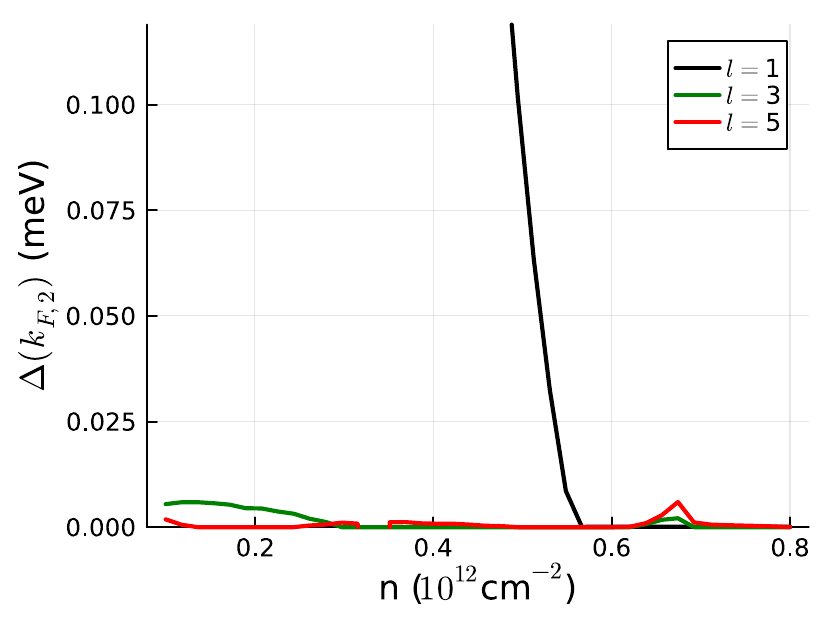}
    \end{tabular}
    \caption{Zero-temperature pairing potentials for $p$-wave (l = 1), $f$-wave (l = 3), and $h$-wave (l = 5) for displacement fields (a) $D = 30\, {\rm meV}$ and (b) $D = 60\, {\rm meV}$. This calculation does not include form factors.}
    \label{fig:supp-potentials-l135}
\end{figure}

We investigate the higher angular momentum channels from our mechanism. Fig.~\ref{fig:supp-potentials-l135} shows the zero temperature pairing potentials at the outer Fermi surface $k_{F,2}$ for $l = 1, 3$, and $5$, corresponding to $p$-, $f$-, and $h$-wave pairing. The pairing potetials are of the order of $10^{-3} \, {\rm meV}$, about two order of magnitude smaller than the typical pairing potential of $p$-wave pairing. A similar result hold for the critical temperature, where we found maximal $T_c$ of the order of $20 {\rm mK}$, where the pairing state satisfies $k_B T_c / \Delta(k_{F,2} \approx 2$. In principle, $f$-wave and $h$-wave pairing are present at large density where $p$-wave is absent, however the corresponding critical temperatures and superconducting gaps are so small that these states are likely not observable.

\section{Pairing with bare Coulomb repulsion}
\label{app:bare_coulomb}

\begin{figure}
    \centering
    \begin{tabular}{ll}
         (a)  & (b) \\
         \includegraphics[width=0.35\linewidth]{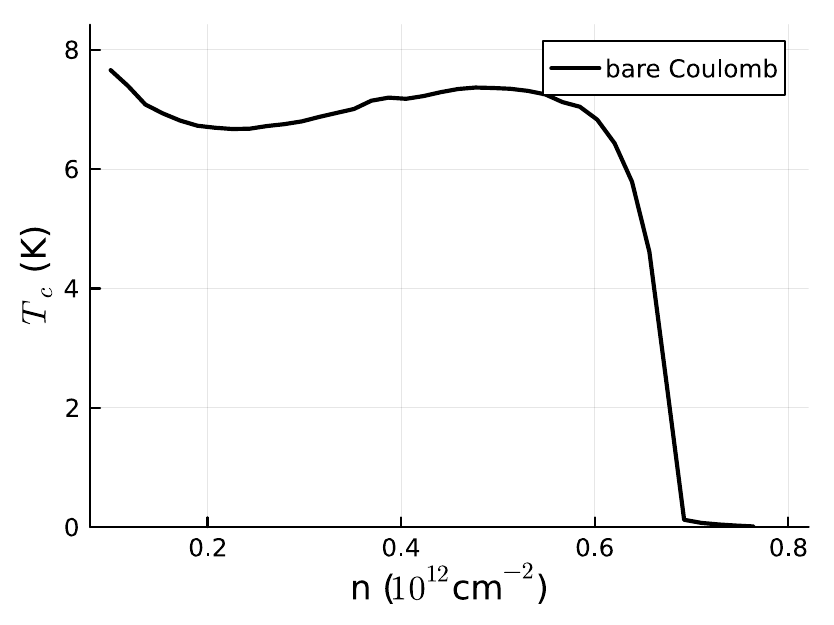} & \includegraphics[width=0.35\linewidth]{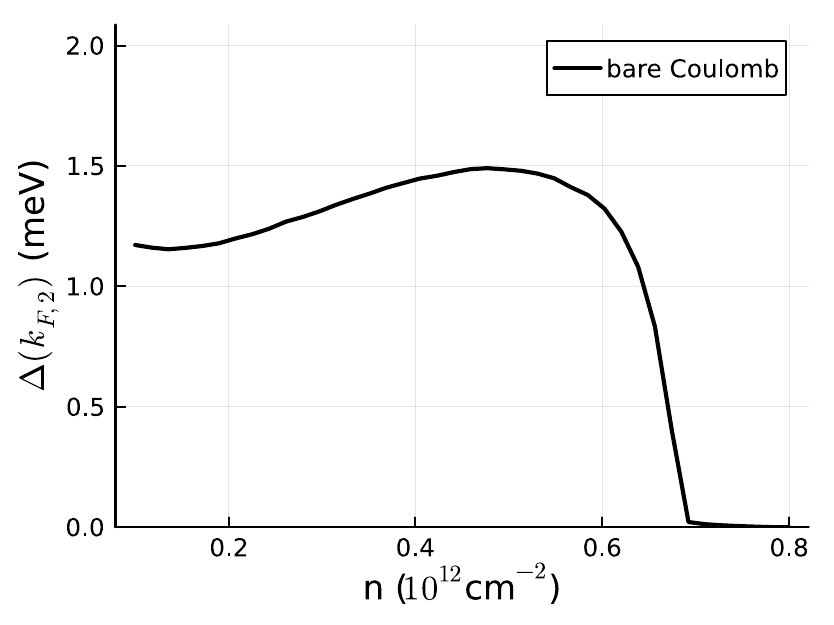}
    \end{tabular}
    \caption{Critical temperature (a) and zero-temperature pairing potential at the outer Fermi surface $k_{F,2}$ (b) at $D = 60\, {\rm meV}$ for bare Coulomb repulsion ($r_K = 0$). This calculation does not include form factors.}
    \label{fig:supp-bare-Coulomb}
\end{figure}

With bare Coulomb repulsion, the maximal obtained $T_c$ is around $8 \ {\rm K}$, see Fig.~\ref{fig:supp-bare-Coulomb}. The strong dependence of $T_c$ on the dielectric screening parameterized by the Keldysh parameter $r_K$ (compare to Fig.~\ref{fig:3} in the main text) indicates that a more detailed modeling of the electron-electron interaction on distances shorter than the thickness of the rhombohedral graphene film is necessary for an accurate prediction of $T_c$. 

\end{appendix}

\end{document}